\documentclass[preprint,12pt]{elsarticle}

\usepackage{amssymb}
\usepackage{amsmath}
\PassOptionsToPackage{hyphens}{url}\usepackage{hyperref}

\journal{Astroparticle Physics}

\begin{document}

\begin{frontmatter}

\title{Deep learning with photosensor timing information as a background rejection method for the Cherenkov Telescope Array}


\author[oxford]{S. Spencer}
\author[oxford,mars]{T. Armstrong}
\author[oxford,heid,desy]{J. Watson}
\author[ciemat]{S. Mangano}
\author[geneva]{Y. Renier}
\author[oxford]{G. Cotter}
\address[oxford]{Department of Physics, Denys Wilkinson Building, University of Oxford, Keble Road, Oxford, OX1 3RH, United Kingdom}
\address[mars]{Centre de Physique des Particules de Marseille (CPPM), Aix-Marseille Universite, CNRS/IN2P3, Marseille, France}
\address[heid]{Max-Planck-Institut f{\"u}r Kernphysik, P.O. Box 103980, 69029 Heidelberg, Germany }
\address[ciemat]{CIEMAT, Avda. Complutense, 40 - 28040 Madrid, Spain }
\address[geneva]{DPNC - Universit\'{e} de Gen\`{e}ve, 24 Quai Ernest Ansermet, CH-1211 Gen\`{e}ve, Switzerland}
\address[desy]{DESY,Platanenallee 6, 15738 Zeuthen, Germany}
\begin{abstract}
New deep learning techniques present promising new analysis methods for Imaging Atmospheric Cherenkov Telescopes (IACTs) such as the upcoming Cherenkov Telescope Array (CTA). In particular, the use of Convolutional Neural Networks (CNNs) could provide a direct event classification method that uses the entire information contained within the Cherenkov shower image, bypassing the need to Hillas parameterise the image and allowing fast processing of the data. 

Existing work in this field has utilised images of the integrated charge from IACT camera photomultipliers, however the majority of current and upcoming generation IACT cameras have the capacity to read out the entire photosensor waveform following a trigger. As the arrival times of Cherenkov photons from Extensive Air Showers (EAS) at the camera plane are dependent upon the altitude of their emission and the impact distance from the telescope, these waveforms contain information potentially useful for IACT event classification. 

In this test-of-concept simulation study, we investigate the potential for using these camera pixel waveforms with new deep learning techniques as a background rejection method, against both proton and electron induced EAS. We find that a means of utilising their information is to create a set of seven additional 2-dimensional pixel maps of waveform parameters, to be fed into the machine learning algorithm along with the integrated charge image. Whilst we ultimately find that the only classification power against electrons is based upon event direction, methods based upon timing information appear to out-perform similar charge based methods for gamma/hadron separation. We also review existing methods of event classifications using a combination of deep learning and timing information in other astroparticle physics experiments.
\end{abstract}

\begin{keyword}
CTA \sep CHEC \sep Deep Learning \sep IACT \sep Background Rejection


\end{keyword}

\end{frontmatter}


\section{Introduction} \label{Introduction}

Imaging Atmospheric Cherenkov Telescopes (IACTs) are a mature technology for the indirect observation of astrophysical very-high-energy (VHE, over ~20 GeV) $\gamma$-rays \cite{tevreview}. There are currently three major IACT arrays that have been operating for roughly two decades: H.E.S.S. \cite{hesscrab} in Namibia, VERITAS \cite{jamieveritas} in Arizona and MAGIC \cite{magiccrab} on La Palma. The single telescope FACT \cite{fact} also operates as a blazar monitor. IACTs operate by observing the Cherenkov light emitted when a VHE $\gamma$-ray undergoes pair production in the atmosphere, creating an Extensive Air Shower (EAS). In order to improve the reconstruction of the energy and direction of the incident $\gamma$-ray, data from an array of such telescopes is combined stereoscopically. Over the past decade these instruments have produced a wealth of new results and have demonstrated the validity of the IACT technique \cite{tevreview}. 

The Cherenkov Telescope Array (CTA) is an ambitious project to build a next-generation VHE $\gamma$-ray facility, which would improve on the sensitivity of the current generation instruments by roughly an order of magnitude \cite{scienceCTA}. The CTA consortium, involved in directing CTA Observatory science goals and array design, consists of over 1400 scientists from 31 countries around the globe. Given the large energy range (20 GeV to 300 TeV) that CTA will operate  over \cite{scienceCTA}, three classes of IACT are required. These are designated as the Small-, Medium- and Large-Sized Telescopes (SSTs, MSTs and LSTs). The 23-meter diameter LSTs are designed for high sensitivity observations of $\gamma$-rays over the 20 GeV-150 GeV energy range, increasing the overlap in sensitivity with space based missions such as the Fermi $\gamma$-ray Space Telescope \cite{Fermi}. The SSTs will be the smallest (approximately 4 m diameter) but most numerous telescopes, and will be spread out over a large ($\sim$4 km$^2$) area in order to maximise CTA's effective area for $\gamma$-ray energies over 5 TeV. The MSTs will provide unprecedented sensitivity to cosmic $\gamma$-ray fluxes in the intermediate energy range. The baseline design of CTA consists of 99 telescopes (4 LSTs, 25 MSTs, 70 SSTs) placed on a southern site at Cerro Paranal in Chile, whereas a smaller array of 19 telescopes (4 LSTs, 15 MSTs, and no SSTs) will be placed on a northern site on Spanish Canary Island of La Palma. In this way, the CTA Observatory will cover the whole sky \cite{scienceCTA}.

This paper is primarily concerned with the SSTs, and in particular we investigate the \textit{Compact High Energy Camera} (CHEC) which is one of the SST prototype cameras \cite{checzorn}. CHEC is designed to work with SST structures that have a dual-mirror (Swartzschild-Couder) optical design. This design allows for a compact camera (approximately 0.4 m diameter) with small scale photosensors (6/7 mm, corresponding to roughly a third of the plate scale of a comparable single mirror design) and a wide field of view ($\sim$0.2 degrees per pixel). There are two prototype telescope designs that CHEC cameras can be attached to: the French-led \textit{Gamma Cherenkov Telescope} (GCT), and the Italian-led \textit{Astrofisica con Specchi a Tecnologia Replicante Italiana} (ASTRI) instrument. In June 2019, it was decided that the final CTA-SST design will be based on ASTRI and CHEC, taking into account the experience gained from all SST designs. Two operational CHEC prototypes have already been built, the most recent of which (CHEC-S) is currently undergoing testing at the Max Planck Institute for Nuclear Physics in Heidelberg, and has recently been used on an observing campaign on the ASTRI prototype. Further details regarding the science performance of the SST sub-array and the complete CTA instrument can be found in Maier et. al \cite{gernotCTA} and Hassan et. al \cite{tarekCTA}.

Before addressing substantially the novel elements of this paper, it is necessary for us to introduce a number of basic 
concepts and terms. \textbf{Supervised Machine Learning} refers to the areas of automated pattern recognition whereby labelled \textbf{Training Data} exists such that a configurable function can be optimised to map an input to a desired output. In our case this function takes the form of a \textbf{Neural Network}, which consists of layers of \textbf{Artificial Neurons} that evaluate a set of inputs (with associated configurable weights and biases) against output \textbf{Activation Functions}.  \textbf{Convolutional Neural Networks (CNNs)} are a variant of neural network primarily designed for use with image data, which can exploit both the translational invariance and relative sparsity of useful information in images to effectively handle such data. \textbf{Recurrent Neural Networks (RNNs)} are specialised neural networks designed for handling data series, and have features such as loops in the network architecture specifically tailored for such use. \textbf{Long Short-Term Memory Networks (LSTMs)} are a particular variant of RNNs that achieve good performance on computer science datasets \cite{Shilon}.  \textbf{Training} describes the process by which the weights in a supervised machine learning algorithm are manipulated such as to minimise a loss function against the labels provided for training data, \textbf{Testing} describes the process of determining these trained networks' efficacy against previously unseen data. Finally, \textbf{Graphics Processing Units (GPUs)} are currently the optimal hardware for performing most deep learning analyses in physics. Whilst a detailed explanation of the algorithmic details of these methods is beyond the scope of this work, we refer the interested reader to \cite{goodfellow2016deep} \cite{erdmannwhite} \cite{dcnn} for further information.

We present in this paper the first work to consider applications of 2-Dimensional Convolutional Long Short-Term Memory (ConvLSTM2D) Networks techniques in the high energy range probed by the SSTs. The significance of this approach lies in the fact that not all EAS are caused by $\gamma$-ray photons, and background rejection is a key component in IACT sensitivity \cite{Shilon}. For most energies, EAS from incident charged hadrons (most commonly protons) outnumber those caused by photons by a factor of roughly 10,000 \cite{Benbow}, though use of trigger selection and parameter-based cuts can reduce this ratio to 1:1 for a bright source like the Crab Nebula using a current generation instrument like H.E.S.S. \cite{Berge07}. However, since the quarks contained within protons experience the strong nuclear force, the typical interactions (such as gluon-gluon interactions, pion production and muon production) they undergo on entering the atmosphere are different to those of photons.  The pions produced travel away from the shower core before decaying, some of which produce $\gamma$-rays which can themselves undergo pair production creating electromagnetic sub-showers of the hadronic shower core. These effects combine, resulting in a wider, less structured EAS compared to a $\gamma$-ray induced shower. Electrons can also be a significant source of background at low TeV energies as they are difficult to distinguish from $\gamma$-rays and undergo similar interactions. There are two small differences between electron and $\gamma$-induced air showers. The first is the process by which the primary particles interact on hitting the atmosphere, a primary $\gamma$-ray will typically loose all its energy in a single pair production event, whereby a primary electron of the same initial energy can loose energy via production of numerous lower-energy Bremsstrahlung photons which can create electromagnetic sub-showers. This can create a relatively greater number of Cherenkov photons at higher altitudes for electron-induced showers. The other small difference between the two is the average altitudes at which they first interact \cite{Sitarek1i}, which is due to the radiative length of an electron being smaller than the pair production length of a $\gamma$-ray. As a result, a primary cosmic-ray electron of the same energy as a primary $\gamma$-ray will begin interacting higher in the atmosphere, resulting in a higher-altitude shower maximum \cite{lypova}.  This work is the first known attempt to determine if new deep learning techniques and new prototype instruments have any potential to aid in discriminating against cosmic-ray electrons using a combination of complete images and photosensor waveform data, suggested as a possibility by Nieto et. al in \cite{nieto2017exploring}. Previous (largely unsuccessful) attempts at the highly challenging task of $\gamma$-electron event separation for IACTs have relied upon indirect estimation of first interaction height from IACT images using either shower image shape or combined shower energy and maximum estimation \cite{Sitarek1i}.

Hillas Parameters form the basis of the observables used for event discrimination in the current generation of IACTs \cite{hillasparams}. After the IACT camera images (constructed from the integrated charge for each camera pixel) have been cleaned using a two-level tailcut technique (the implementations of which vary by telescope, see for example \cite{Benbow}\cite{magictime}), these Hillas Parameters are obtained \cite{weekestev} from the second moments of the image. The Hillas Parameters relevant for event classification are the width and length of the ellipse which characterize the longitudinal and lateral development of the shower. Initially Hillas Parameters were simply used to perform data cuts to separate hadronic showers from $\gamma$-ray induced showers based on their differing morphology. However in recent years, more sophisticated methods using Boosted Decision Trees (BDTs) or Random Forests (RFs) eventually became the preferred methods for incident particle classification \cite{hessbdt}. These rely upon parameterised data such as the Hillas width and length, along with total integrated charge and other similar parameters derived from Monte-Carlo simulations. 

Hillas parameter based techniques don't however take advantage of the full camera image of the EAS, and as such, subtle details (such as hadronic `halos' \cite{model++}) in the images are not taken into account. This becomes an issue at the sensitivity boundaries that CTA is aiming to considerably improve (particularly in the case of weak or very extended sources), as hadronic and electron induced showers can closely resemble those generated by gamma-rays. This motivates us to investigate new analysis techniques for event discrimination in order to improve the IACT sensitivity.

Template analyses \cite{cat}\cite{3danalysis}\cite{model++}\cite{impact}, which benefit from the fine pixellisation of IACT cameras, form the current main alternative to Hillas parameterisation. They rely on the generation of a library of template IACT images, constructed using either full Monte-Carlo simulations or semi-analytic models. The images from the IACTs are then compared to the templates via a pixel-by-pixel minimisation process, such that the template most resembling the IACT image can be found and thus the shower properties inferred. In particular, such techniques have been commonly applied for directional reconstruction (such as ImPACT \cite{impact}), although the H.E.S.S. Model++ analysis chain uses fitting against semi-analytic models of gamma-ray showers as a background rejection method \cite{model++}.

We split the main body of our paper into five sections. In section 2, we discuss recent advances in machine learning techniques in astroparticle physics, particularly concerning the use of timing information in combination with deep learning techniques as a background rejection method. In section 3 we detail the simulated datasets we use and in section 4 we describe the analysis we perform. In section 5 we present our results and we conclude in section 6. A preliminary version of this work was presented in \cite{samicrc}.

\section{Background and Related Work} \label{RelatedWork}
New supervised machine learning methods such as CNNs have seen wide investigation across the physical sciences in recent years \cite{Shilon} \cite{ErdmannAuger}\cite{erdmannwhite}\cite{tunka}. This has been driven by a number of factors: improvements in computing power, improvements in algorithmic development, and by major investment, utilisation and innovation from industry \cite{dcnn}. In this section, we review recent relevant use cases of these new machine learning methods in astroparticle physics.

Shilon et al. \cite{Shilon} recently applied deep learning based $\gamma$-hadron separation methods to H.E.S.S. data . In their work, they feed images from the CT1-4 telescopes into convolutional layers that then feed into an LSTM Network. This technique, called the Convolutional Recurrent Neural Network (CRNN) method, treats the set of four charge images from the four IACTs as a time series. The authors of the Shilon et al. work justify this network architecture using the assumption that that the images with the largest size parameter (total charge) are those closest to the shower core, and that this ordering compensates for the lack of available temporal information in their case \cite{Shilon}. Using this, they observed performance in background rejection superior to the current BDT paradigm with simulations, and were able to perform a significant detection of PKS 2155-304 with real data. The CRNN technique achieved a 50.8$\sigma$ detection with this data relative to a BDT which achieved 46.1$\sigma$, though the statistical significance of this difference has not been completely explored. In particular, this source is extremely bright, and so this data is likely $\gamma$-dominated. Whilst this network architecture allowed for the first CNN-based detection of an astrophysical source, the Shilon et al. work's authors' assumption that the ordering of the images in the LSTM is significant has not necessarily held up to subsequent scrutiny, with other charge image orderings \cite{ariconf} being seen to provide equivalent or superior performance. As a result, for the purposes of our work in later sections, we treat the LSTM-type features in similar networks as a computational tool. Whilst this tool is useful for avoiding the need to stack images from multiple telescopes into single images in order to apply CNN-type analyses (which had been the standard previous to the Shilon et al. work and doesn't scale well in the case of multiple IACTs \cite{Shilon}\cite{salvatore} as details in images from high multiplicity events is lost), we ultimately don't assume the use of the LSTM features in such networks to be of further physical significance.

It is not currently clear how well these CNN-based event classification methods will work with dimmer sources given their high level of sensitivity to small differences between real data and simulations. We briefly explore such issues related to real data further in section \ref{Conclusions}. A key advantage such deep learning methods is the speed at which they can perform classifications of complete IACT images ($\sim$1 kHz for Mono-telescope analysis using a NVidia 1080Ti GPU), which is a major factor to consider given the high expected array trigger rate of CTA ($\sim$10 kHz \cite{trigrate}). However, such methods require training beforehand and this can be computationally intensive.

The use of EAS timing information for event classification in IACT data has been considered as early as in \cite{Paulathesis}. More recently, it has been used as an image cleaning method on MAGIC (whereby pixels associated with Cherenkov light are partly selected based upon photon arrival time) \cite{magictime}, prior to their RF analysis. Similar studies to this have also been performed with VERITAS \cite{jamietime}. MAGIC also use parametric timing data such as a time gradient (how fast the arrival time changes along the major IACT image axis) and Time RMS (the RMS arrival time of photons to all pixels that survive an image cleaning process) as part of their RF-based background rejection method \cite{supermagictime}. In that work \cite{supermagictime}, the MAGIC collaboration also demonstrate the dependency of timing structure upon the impact parameter of the shower, also information useful for background rejection. The attraction of deep learning in our use case is the ability to use this timing information for background rejection directly, in addition to the other pixel-wise data (such as evidence of electromagnetic shower sub-structure in hadronic showers), without the need for parameterisation. An illustration of this information can be seen in Figure \ref{fig:wfplot}. Whilst this is the first work to attempt such a combination of timing information and deep learning for IACTs, this is not the case for other air shower experiments, and so in the remainder of this section we investigate the approaches taken by other groups.

Erdmann et al. \cite{aug1} investigated the use of deep learning to perform energy and angular event reconstruction of air showers for a model hexagonal array of 81 ground based water Cherenkov detectors using a custom simulation code. They feed in waveforms from their array into a CNN, which is concatenated with 2D feature maps of the first particle arrival time and integrated charge for the entire detector grid approximately half way through their network architecture. These feature maps are then treated with separable convolutions, as it is not expected that the correlations between the feature maps will be coupled to spatial correlations within the maps themselves. The use of this information allows them to achieve an energy resolution of 5\% in the EeV energy range, compared to 8\% without the additional concatenated feature maps. We do not attempt a similar approach here (as has recently been attempted as an expansion to the Shilon et al. work by \cite{ParsonsOhm}) as it becomes even more difficult to interpret the behaviour of the neural network when parameterised data (such as impact parameters) is mixed with image data. In a later paper \cite{ErdmannAuger}, the authors investigate their analysis robustness to real data effects by creating air shower simulation datasets with different divisions of energy between the muonic and electromagnetic components of the showers. They successfully train a 1-dimensional Generative Adversarial Network (GAN) in order to attempt to refine their 1-dimensional waveform traces from one simulation set such that it resembles the second set of simulations more closely. This is an attempt to handle the problem that minute differences between training and test sets in machine learning analyses can cause discrepancies; however they find that the differences between their simulated datasets are insufficiently large to affect their quality of the energy reconstruction. However, 2D image data is substantially more complex and it is unclear how well a GAN would perform in our case. A study has also been performed on data from the Tunka-Rex experiment \cite{tunka} on 1-dimensional timing information regarding the radio emission from $>100$ PeV air showers which provided improved background suppression, however this was only performed on single traces from a single antenna and not the complete array.

In \cite{icecube1}, the IceCube collaboration perform reconstruction of muon neutrino events using deep learning. Each of the 5160 in-ice digital optical modules (DOMs, a photomuliplier with associated electronics enclosed in a sealed unit) records a waveform when Cherenkov light emitted from the particles generated by interactions in the ice reaches the photomultiplier in the DOM. In that work, they parameterise the registered waveforms using 7 variables (sum of all charges, number of pulses, time of first pulse, time of last pulse, the average time of a pulse, the standard deviation of pulse times and the highest pulse charge) which are then inserted into a 3-dimensional neural network. On simulated events, they are able to obtain an improvement in energy resolution relative to a more conventional reconstruction method. This approach to parameterizing waveforms partially inspired our work, as use of full CHEC waveforms with e.g. 3D CNNs is likely to be prohibitively computationally expensive, especially given the fact that we work with arrays consisting of multiple telescopes. Choma et al. \cite{icecubegraph} also investigate peak waveform timing information as an input to their novel graph-network based architecture (which takes graph objects with edges and vertices rather than arrays as inputs), which they use to classify IceCube events. However, the particular network architecture used in this work is currently not suitable for use on images from IACTs.

\section{Datasets} \label{Datasets}

All current IACT cameras have the ability to record some measure of timing information. Given that many of the camera prototypes under development for CTA (and many other current generation cameras) have the ability to read out the entirety of the waveforms from their photosensors, we are motivated to explore how this complete waveform information could be used for event classification. In particular, we seek to find a way to use this extra information in combination with new deep learning techniques in order to take maximal advantage of the information available about the EAS as possible. In an attempt to keep our exploration of this challenge tractable, we restrict ourselves to differential consideration of event classification alone. This is only one particular element of a complete IACT analysis necessary to perform observations of real sources \cite{Berge07}\cite{LiMa}.

To investigate this task, we require large, labelled training datasets. In our case, this was Monte Carlo simulations of $\gamma$-ray, hadronic and electron induced EAS. To generate the simulated datasets, we used the CORSIKA package (version 6.990) \cite{corsika} to simulate the particle physics of the EAS and simulate the Cherenkov light they produce. We then passed this data to the sim\_telarray (version 2017-09-01) package \cite{BERNLOHR} to simulate the telescope optics and camera electronics using the CTA Prod3b array model \cite{prod3b}. Though it should be noted that these simulation models (along with the associated \textit{ctapipe} based analysis pipeline) are under continuous refinement. Parameters describing the CORSIKA simulations are shown in Table \ref{table:Datasetparams}. For protons, only around a third of the energy of the incident particle goes into producing the electromagnetic component which ultimately produces Cherenkov light. As such, the hadronic showers that are a background to our IACT telescopes have higher primary energies than the $\gamma$-rays with the same total Cherenkov signal, and this is reflected in our simulations. We simulated 4 GCTs \cite{gct} equipped with CHEC-S cameras \footnote{In this work, we simulate a 96 ns camera readout window with 96 samples.}, although the conclusions of this work are relevant to other instruments with similar capabilities (such as CHEC-S on ASTRI). The telescopes are arranged in a cross-configuration (+) on the Cerro Paranal site separated from the centre of the array by +-80 meters, and are simulated in the standard parallel pointing mode. This array configuration was chosen both to resemble the one in \cite{Shilon} and also to produce data that could be processed with finite GPU resources. Particularly in the case of the SSTs, larger simulated arrays will eventually be required for detailed studies into the use of CNNs with VHE events that trigger readout from many cameras.

Despite a growing amount of literature and work on the use of CNN-based classifiers for IACT background rejection \cite{Shilon}\cite{samicrc}\cite{dl1dh}, there is currently little consensus or detailed study of the types of datasets required to train such networks (especially in the case of application to real data). Even the pioneering Shilon et al. work used pre-existing simulations to avoid the significant computational cost of investigating this \cite{Shilon}. CNNs are potentially sensitive to different features in image data compared to template fitting methods, and the fact that they are capable of handling complete images rather than parameterised data means that the approaches used in most current analyses cannot be assumed to be optimal with CNN-based classifiers. As a result, we investigate two scenarios. In the first, we attempt to differentiate between simulated $\gamma$-rays from a point source against a diffuse, simulated background of proton and electron events. For the purposes of this work we define point source simulations as those with a CORSIKA opening angle of $0^{\circ}$, and diffuse simulations as those with a CORSIKA opening angle larger than $0^{\circ}$. Whilst using a simulations approach similar to this for observations would mean that generating new simulations and performing network training would be required for every run ($\sim$30 minutes of observing time), it is our opinion that this is the most promising approach for applications of CNN-based classifiers to real data in the medium term. This is as discrepancies between these simulations and real data are a significant issue for such sensitive techniques. We believe that a run-wise approach, which is now being used by H.E.S.S. \cite{rws} (and is similar to this point source run) might possibly help in observing real sources. We explore this further in section \ref{Conclusions}. Additionally, we postulate that the high sensitivity of CNN-type methods to pixel wise features may allow for improved background rejection power (and ultimately improved source significance) using a combination of directional and morphological pixel wise information relative to more conventional directional and shape Hillas parameter cuts. However, such a run wise approach comes with significant caveats, and is not suitable for use in all observing circumstances. For example, in using this approach blind surveys would be prohibitively computationally expensive, there would be substantial challenges related to source confusion in the galactic plane, and it would not be possible to observe extended sources (or fields of view with multiple sources). Realistically, CNN-type background rejection methods will require further study in the simplest analysis case of strong, isolated point sources using real data, before considering more complex analyses. In the second scenario, we attempt to discriminate between diffuse $\gamma$-ray events, diffuse proton events and diffuse electron events, this is similar to the simulation set-up in the Shilon et. al work \cite{Shilon}. This indicates the classification power based upon shower morphology alone and also allows us to test the sensitivity of the methods in detecting the direction of the shower by comparison to the point source results. Given the high computational cost these methods entail, both in terms of GPU training time and generation of sufficient Monte-Carlo simulations, we are investigating these methods as a long term analysis prospect, and not as an analysis method suitable for immediate use for CTA. This is under the assumption that available computing power will increase with time.
\begin{table}[ht]
\centering
\resizebox{\textwidth}{!}{
\begin{tabular}{r|c|c|c|c}
    \textbf{Parameter} &\textbf{Point Source $\gamma$} & \textbf{Diffuse $\gamma$} & \textbf{Diffuse p} &\textbf{Diffuse e}\\
    \hline
    No. Training Events Point Source Run& 360995 &-&361135 &361135\\
    No. Testing Events Point Source Run& 239128&-&239227&239227\\
    No. Training Events Diffuse Run& - &365763&365903&365903\\
    No. Testing Events Diffuse Run& -&240453&240552&239894\\
    Energy Range (TeV) &0.3-330&0.3-330&1-600&0.3-330\\
    Opening Angle &$0^{\circ}$&$0^{\circ}$-$10^{\circ}$&$0^{\circ}$-$10^{\circ}$&$0^{\circ}$-$10^{\circ}$\\
    Spectral Index & -2 & -2 & -2 & -2\\
    Zenith Angle &$20^{\circ}$&$20^{\circ}$&$20^{\circ}$&$20^{\circ}$\\
    Azimuth Angle &$0^{\circ}$&$0^{\circ}$&$0^{\circ}$&$0^{\circ}$\\
    Cherenkov light waveband (nm) &240-700&240-700&240-700&240-700\\
\end{tabular}}
\caption{Parameters used in our two datasets. In the point source run, the point source $\gamma$-rays are mixed in equal ratios with the diffuse proton and diffuse electron events. In the diffuse run, all three event classes are diffuse.}
\label{table:Datasetparams}
\end{table}

The data from the event files generated by CORSIKA/sim\_telarray was then fed into a python script to process them into calibrated waveforms in units of photoelectrons (p.e.) using the prototype \textit{ctapipe} (v0.6.1) software package \cite{ctapipe}\cite{ctapipe2}. This process includes pedestal subtraction and smoothing with a Hanning filter \cite{hanning} of length 8. Our analysis code also extracts the waveform parameters (such as the Full Width Half Max), and makes use of the \textit{numba} \cite{numba} (v0.45.1) just in time compilation package such that this parameter extraction can be performed at high speed, making the task of extracting these parameters for a large number of events feasible. The $\gamma$-ray, proton and electron data is then randomly mixed together in approximately equal quantities and then saved in the HDF5 \cite{hdf} file format for convenience and long term storage. The use of roughly the same number of events from each class, whilst the current standard practice with these CNN-type methods \cite{Shilon}, is a simplification to ease network training. Not every $\gamma$-ray source on the sky is equally bright, and so real test data will be imbalanced. In particular, this is complicated by trigger selection. As a result, there is a need for a future detailed investigation into the effects of using more realistic, imbalanced training data with CNN-type IACT background rejection methods, as such imbalanced datasets require specialised deep learning treatments \cite{imbalance}. The data processing, along with the CORSIKA/sim\_telarray simulations was performed on the European Grid Infrastructure (EGI) using the CTA-DiRAC middleware \cite{luisa}. Minimal code to recreate the results in this work can be found, along with trained Keras models, at \url{https://www.github.com/STSpencer/wavelearn\_release}.
\section{Methods} \label{Methods}

To begin the discussion of our methods, we will now define a number of terms related to the preparation and use of supervised deep learning methods for the unfamiliar reader.  An \textbf{Epoch} represents one complete pass over the entire training dataset during training. As GPUs have a finite amount of \textbf{Video Random Access Memory (VRAM)} with which to perform Tensor operations, events are loaded in to the GPU during training in batches smaller than the size of the total dataset. \textbf{Batch Size} is the number of events (each with a single label) in a single batch. \textbf{Hyperparameters} are values (such as the number of layers and convolutional filter sizes) that define the network architecture and learning process and are typically user-selected, though this can have significant performance effects that we discuss later in this section. Further basic machine learning definitions can be found in \ref{MLdefs}.
 
\begin{figure}
  \centering
  \includegraphics[width=0.7\textwidth]{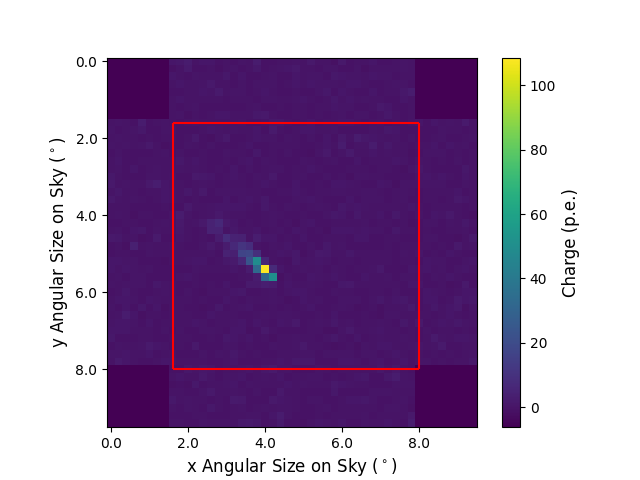}
  \caption{A charge image from a simulated $\gamma$-ray event from the point source run dataset as described in section \ref{Datasets}. The region that survives the geometry cut is highlighted in red, the holes in the camera geometry to accommodate the flasher calibration system can be seen in the corners.}
  \label{fig:gammacut}
\end{figure}

\begin{figure}
  \centering
  \includegraphics[width=0.7
  \textwidth]{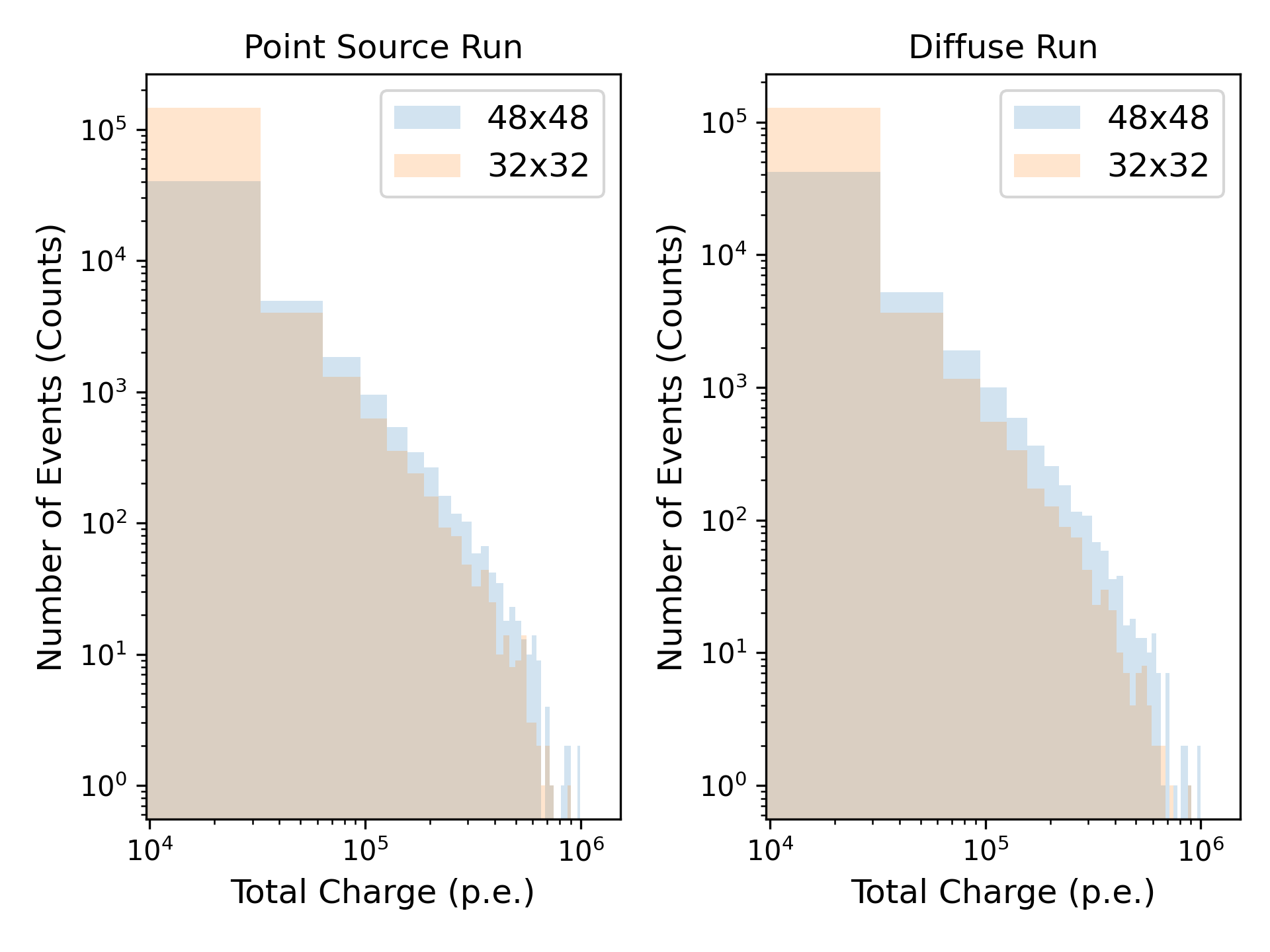}
  \caption{Total integrated charge from all four telescopes over the two datasets used, as described in section \ref{Datasets}. Comparisons between the complete camera (48x48) and the cropped central region (32x32) are shown.  Most of the charge losses appear to be Night Sky Background photons as the two distributions are smooth and close together. The same logarithmic binning is used for both the cropped and uncropped regions.}
  \label{fig:chargehist}
\end{figure}
\begin{figure}
  \centering
  \includegraphics[width=0.7
  \textwidth]{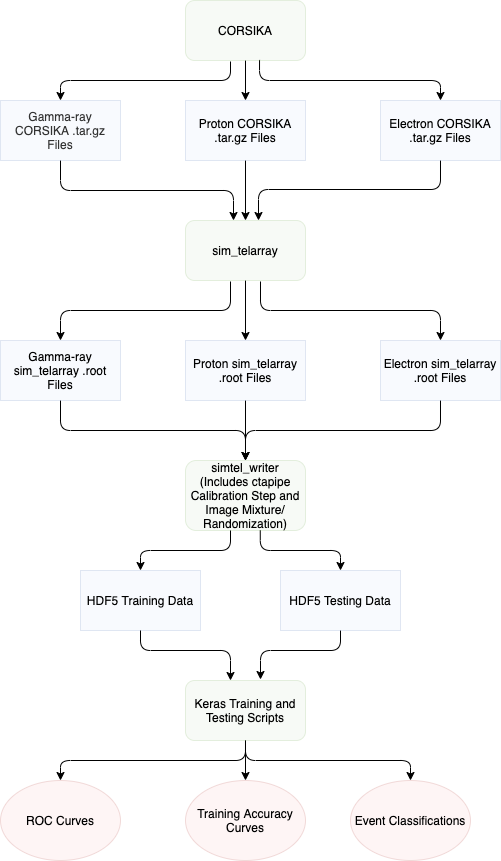}
  \caption{The analysis pipeline used to generate the results in this paper. Ellipses denote analysis products, cornered rectangles denote stored files and rounded rectangles denote code elements.
  }
  \label{fig:wavelearn}
\end{figure}

In order to perform our analysis, we used the \textit{Keras} (v.2.1.5) \cite{Keras} deep learning Python package with the \textit{Tensorflow\_gpu} (1.7.0) backend \cite{tensorflow}. We also used the \textit{scikit-learn} \cite{scikit} (v0.21.3) and \textit{matplotlib} (v.3.1.1) \cite{matplotlib} packages to analyse and plot the results. The analysis pipeline we developed to perform this is illustrated in Figure \ref{fig:wavelearn}. Similarly to \cite{Shilon}, we use a mixture of convolutional layers and recurrence (specifically ConvLSTM2D layers, algorithmic details of which can be found in \cite{shi}) to robustly handle data from multiple telescopes. We compared four different techniques (Methods A-D) for classifying events on a like for like basis with these networks. These are summarised in Table \ref{table:methods} and detailed in the next three paragraphs. 

\begin{table}[ht]
    \centering
    \resizebox{\textwidth}{!}{
    \begin{tabular}{c|c|c|c}
    \textbf{Method} &\textbf{Pixel Maps Used} & \textbf{Ordering} & \textbf{Parameters}\\
    \hline
    A& Timing& Median Peak Time & 293,673\\
    B& Charge, Timing, Mean Amplitude, Peak Amplitude & Median Peak Time & 301,233\\ &FWHM, RMS, RT, FT& \\
    C& Charge & Size Parameter & 293,673\\
    D& Charge & Median Peak Time & 293,673\\
    \end{tabular}
    }
    \caption{Summary of methods used. The ordering of input channels into the ConvLSTM2D networks is as presented in this table. The number of trainable parameters is also shown, highlighting the effect of using a different number of channels in method B.}
    \label{table:methods}
\end{table}
Part of the challenge in \cite{Shilon} was developing an unbiased method to handle Cherenkov camera images that are from a non-square (typically hexagonal) pixel grid \cite{Hexagdly}.  CHEC has largely square pixel geometry, bar for holes in its corners to accommodate the flasher calibration system. In our work, in order to be conservative and minimise the risk of potential geometric bias in which event classifications might be affected by proximity to these corners (which could be the case if they were instead filled by constant values), we simply crop images to the central region of the camera. This can be seen in Figure \ref{fig:gammacut}. This is unlikely to have a serious effect on the signal from the EAS as our simulation set-up (described in section \ref{Datasets}) produces images where the shower is close to the centre of the camera (as shown in Figure \ref{fig:chargehist}). For the same reason, we also neglect in our analysis to take account of small amounts of `dead space' between CHEC photomultiplier blocks. This also does not affect the fundamental conclusions of this work, as they are on a differential basis. More advanced tools are currently under development to cope with these issues \cite{dl1dh} \cite{thomas}.

In method A, we use the available waveform information to generate a 2D timing pixel map across the camera plane. We then select the pixels that have a large waveform amplitude, and set the other pixels in the pixel map to zero, in order to highlight the position of the shower in the image. The location of the peaks themselves is found using a wavelet convolution of width 8 and the \textit{scipy} (v1.4.1) \textit{signal.find\_peaks\_cwt} function \cite{scipy}\cite{findpeaks}. These four timing pixel maps (one for each telescope) are then fed into the network without anything else, in the order of increasing median photon arrival time.

\begin{figure}[ht]
  \centering
  \includegraphics[width=\textwidth]{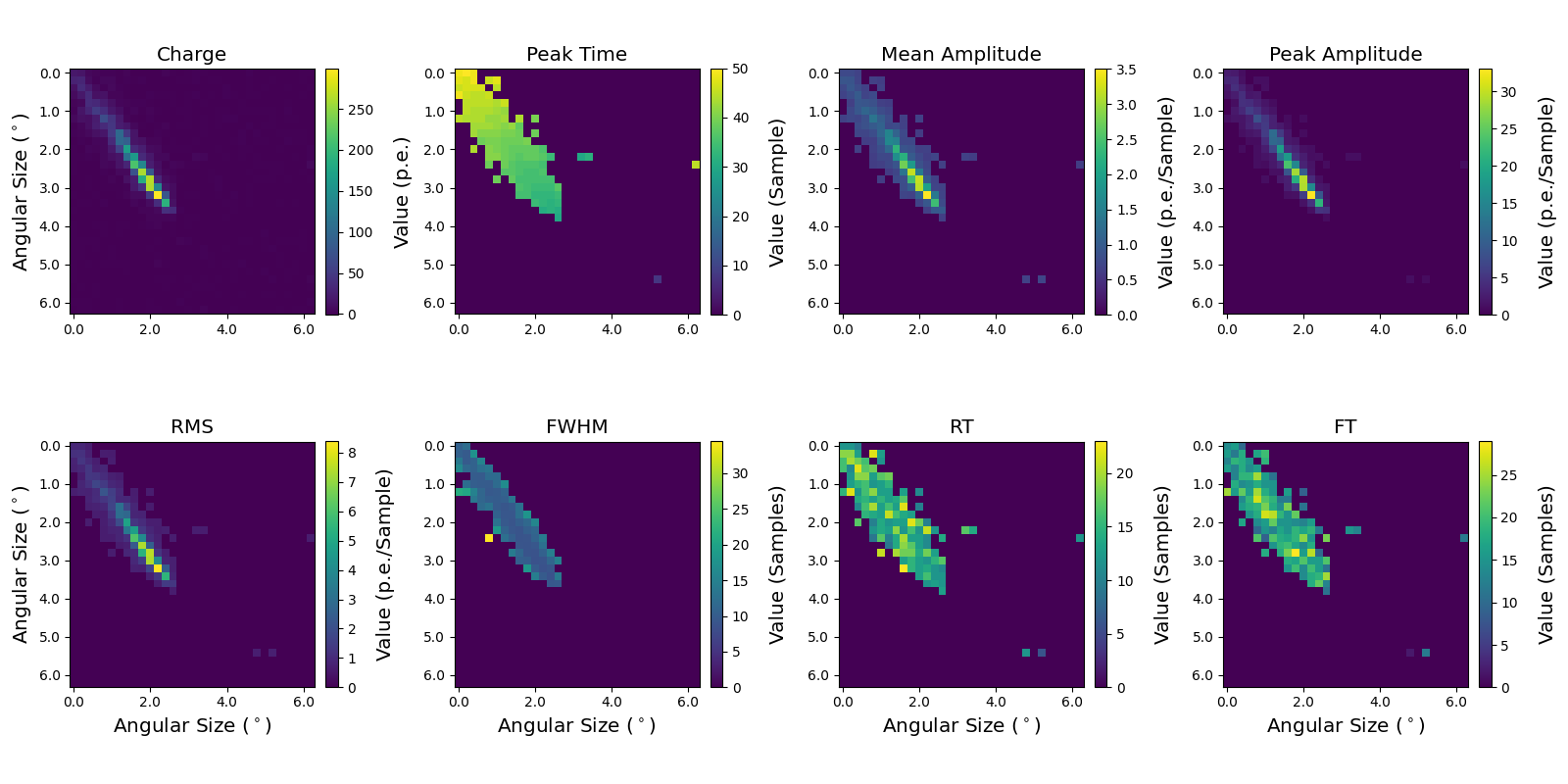}
  \caption{As an illustration of the different pixel maps used with Method A, this figure shows pixel maps used (cut to the central region of the camera) from a 56 TeV $\gamma$-ray event in the point source dataset for one telescope. Note that these pixel maps are normalised prior to analysis, and that 1 sample represents approximately 1ns.}
  \label{fig:wfplot}
\end{figure}

In method B, we feed in the integrated charge image (without pixel selection) as the first channel to the ConvLSTM2D. We then feed in the timing pixel map created using the same technique as Method A as the second channel, and similarly cut 2D pixel maps of six other parameters from the smoothed waveform: the mean amplitude, the peak amplitude, the root mean square value (RMS), the full width at half maximum (FWHM, calculated using a B-spline interpolation \cite{scipy}), the pulse rise time (RT) and the pulse fall time (FT) relative to the peak of the waveform. Both the RT and FT are calculated using a change-in-gradient method. This results in four images each with eight channels being fed into the ConvLSTM2D; examples of these pixel maps are shown in Figure \ref{fig:wfplot}. These channels are treated as extra dimensions in the input tensor, in the same way that Red, Green and Blue channels would be for a colour image.

Method C uses only charge images, ordered such that those with the largest total charge appear first in the LSTM sequence (as in \cite{Shilon}). It should however be noted that we do not perform an exact replication of the method in \cite{Shilon}. This is due to a number of factors. Firstly, we consider different instruments with differing camera designs and operational energy ranges. Secondly, we use a marginally different ConvLSTM2D network architecture (due to the CRNN code from \cite{Shilon} being proprietary), this differs from the CRNN architecture in that recurrent (LSTM-like) features are included from input layer until the final layer of the network (though the basic principle of merging convolutional and recurrent features is the same). We also use differing hyperparameters, partially due to this different architecture, and partly because the differing plate scales and operational energy ranges of the H.E.S.S.-1 cameras and CHEC means that there is no guarantee that the ideal hyperparameter choice will be the same for both instruments. This hyperparameter choice can have a significant effect in CNN-based analysis \cite{hyperopt}. Finally, in order to test the efficacy of CNN-based background rejection against electrons, we include these (realistically) in the analysis whereas \cite{Shilon} did not. This would cause differences with the training of any machine learning classifier (particularly in the diffuse case) as there are two nearly identical data samples with differing class assignments.

In method D, we extract the median photon peak time from the waveforms and use this as a time proxy to order the images in place of the sum of the photomultiplier charges (as used in method C). In this case the only images actually fed to the network are those of integrated charge. As such the only difference between methods C and D is the order in which the charge images for an event are fed into the ConvLSTM2D, such that we can test what effect this has. Therefore methods A, C and D all result in 4 images each with 1 channel (i.e. grayscale) being fed into the ConvLSTM2D network.

\begin{figure}
  \centering
  \includegraphics[width=0.9\textwidth]{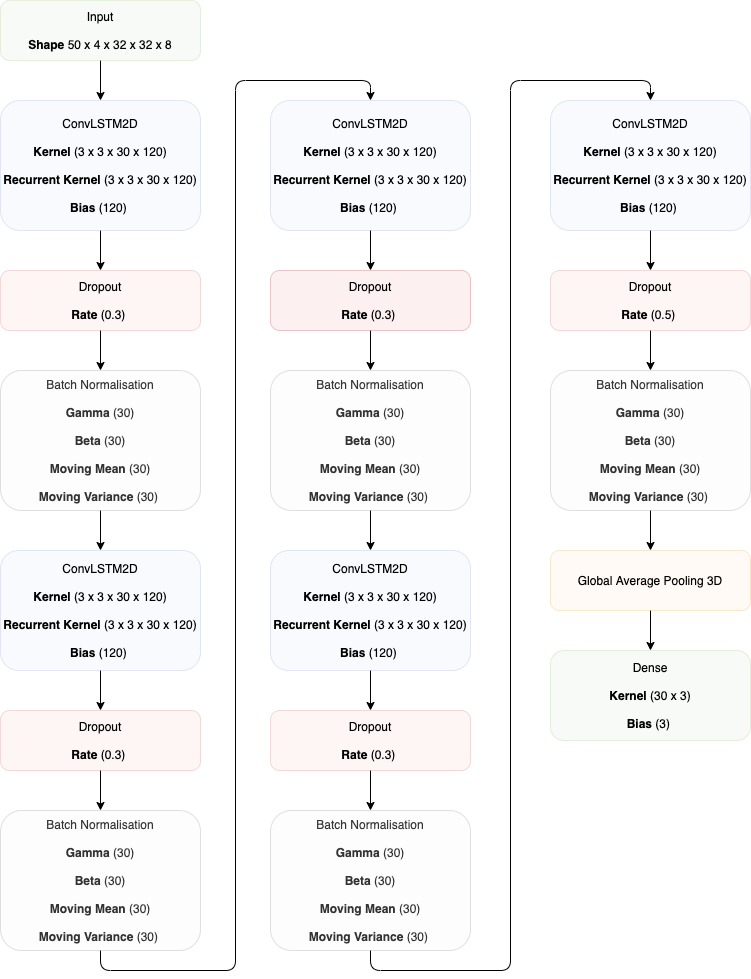}
  \caption{Example of our CNN-based ConvLSTM2D architecture. The only difference between the various methods A-D was the shape of the input vector, as some methods used more pixel maps than others.
  }
  \label{fig:model}
\end{figure}

The datasets used, along with all training hyperparameters (see Table \ref{table:hyperparams}), were the same across all the four methods during the initial training. As this is intended as a differential study, we did not perform hyperparameter optimisation. Whilst there are Bayesian methods such as hyperopt \cite{hyperopt} for performing hyperparameter optimisation with CNN-based classifiers, it would not be trivial to cross-optimise the networks fairly across the four methods described here and it would also involve currently unfeasible computational cost given the size of our datasets. Optimising the networks for say method B might disadvantage Method A unfairly. As such, we believe the least biased option is to use default hyperparameter options across the four methods. The ConvLSTM2D network architecture was also identical across the four methods, with the exception of the number of channels in the input vector (1-8).  This architecture is shown in Figure \ref{fig:model}.

\begin{table}[ht]
    \centering
    \resizebox{0.7\textwidth}{!}{
    \begin{tabular}{c|c}
    \textbf{Hyperparameter} & \textbf{Value}\\
    \hline
    Batch Size & 50 \\
    Training Epochs & 30 \\
    Training Optimiser & Adadelta \cite{adadelta} \\
    Adadelta Learning Rate & 1.0\\
    Adadelta Decay Factor &0.95 \\
    Output Layer Activation Function & Softmax\\
    \end{tabular}  
    }
    \caption{Training hyperparameters used; these were the same throughout all the networks during the initial training for both the point source and diffuse cases. The diffuse ES runs also used these hyperparameters.}
    \label{table:hyperparams}
\end{table}

In order to prevent over-fitting, the ConvLSTM2D layers are followed by dropout and batch normalisation layers, and recurrent L2 regularisation \cite{Keras} is employed on the first two ConvLSTM2D layers. At the end of the network, global average pooling is used to connect to a final dense output layer of size 3 with a softmax activation function. The network is trained for 30 epochs with a batch size of 50 events per step, where the images and pixel maps have amplitudes normalised to the range [0,1]. There is a question over the efficacy of normalisation of IACT images with CNN-based methods as it can potentially prevent generalization, but in this case we observed superior overall training performance when the images were normalised.

\section{Results} \label{Results}
\begin{figure}[t]
  \centering
  \includegraphics[width=\textwidth]{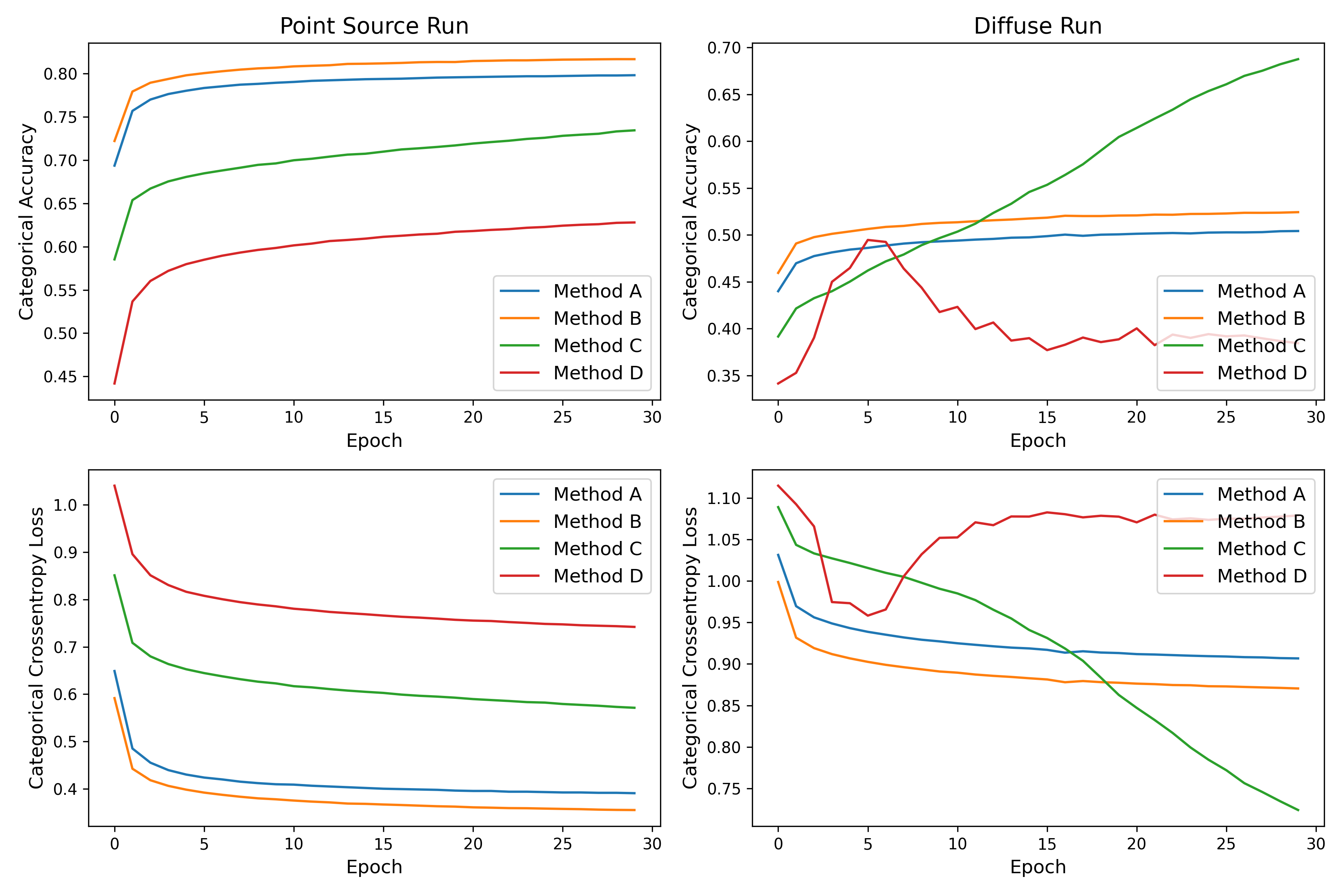}
  \caption{Initial training losses and accuracies for the four methods computed using the training dataset. Reasonable convergence is achieved for most methods given finite computational resources, but methods C and D in the diffuse case show some evidence of overtraining given their poor performance on the test data. It should also be noted that categorical accuracy and AUC values are not identically calculated metrics and the absolute values can differ.
  }
  \label{fig:trainlog}
\end{figure}

\begin{figure}
  \centering
  \includegraphics[width=\textwidth]{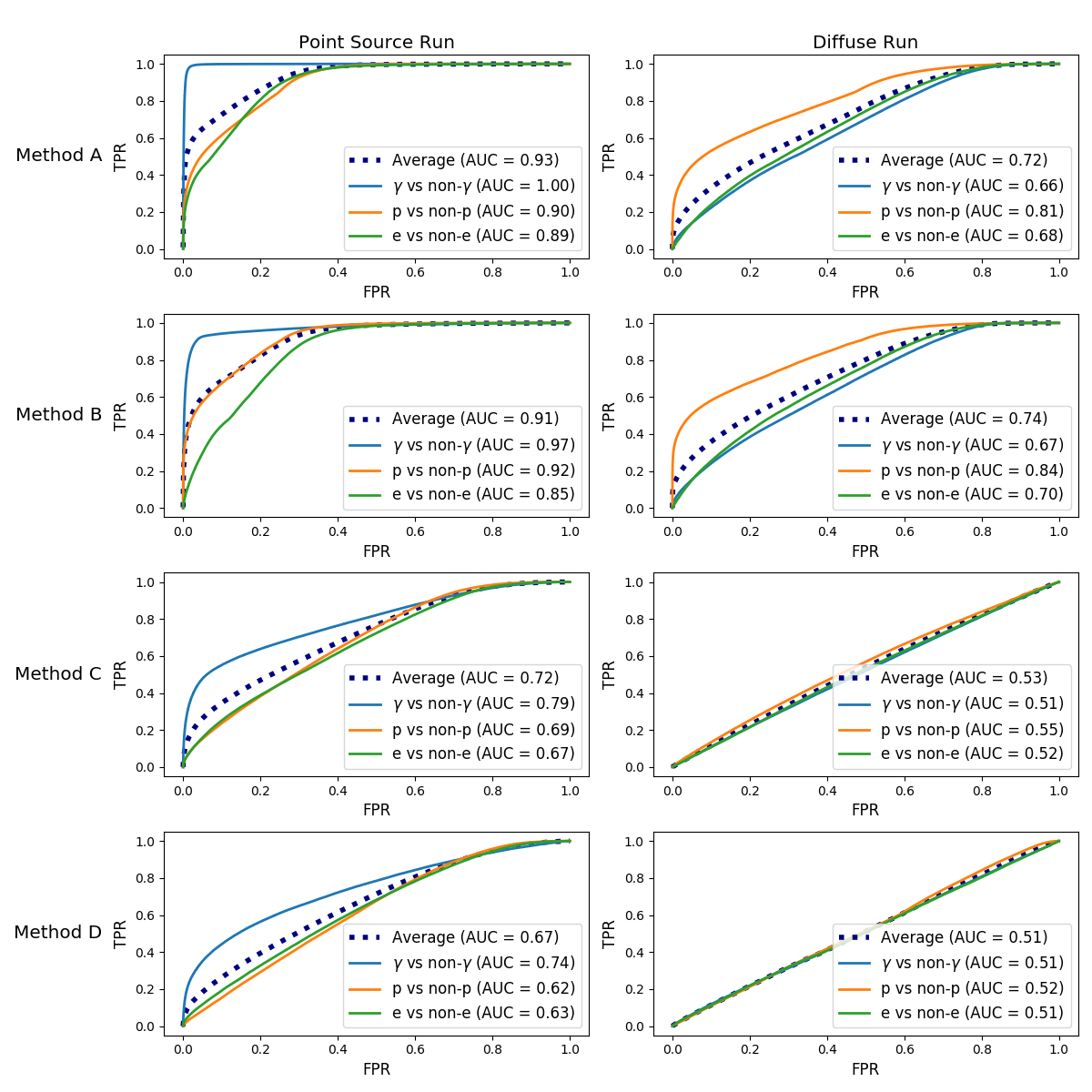}
  \caption{Receiver Operator Characteristic (ROC) curves for the four methods described showing the True Positive Rate (TPR) against the False Positive Rate (FPR). The Area Under Curve (AUC) performance metrics (the integral under these curves) are also shown. Deep learning analyses typically consider one versus all scenarios when calculating ROC curves \cite{scikit}, which is what we display here.  This means we consider if an event is classified a $\gamma$-ray or not a $\gamma$-ray when calculating the FPR and TPR in the multi-class scenario. Macro-averages across the three classes are also shown. It should be noted that the AUC results here and in Figure \ref{fig:ROC2} are only quoted to 2 decimal places and there are $\sim$700,000 testing events. As such, even in the optimal point source case, there are a number of misclassified events (i.e. $\sim$5500 $\gamma$-hadron misclassifications and $\sim$7000 $\gamma$-electron misclassifications for method A).
  }
  \label{fig:ROC}
\end{figure}

\begin{figure}
  \centering
  \includegraphics[width=\textwidth]{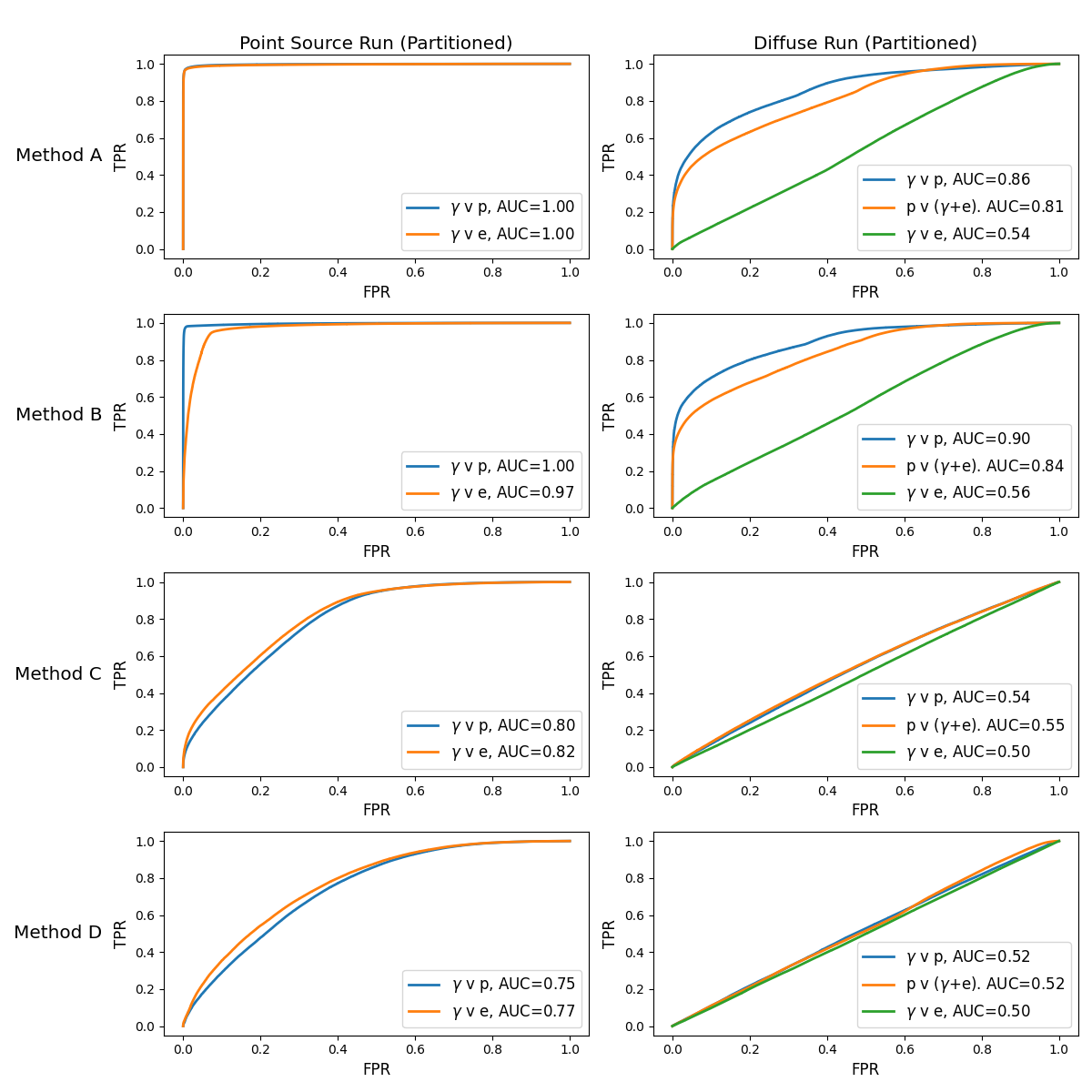}
  \caption{The same results from Figure \ref{fig:ROC}, this time shown in an astrophysical context. In this case we remove the other particle classifications from the analysis and consider only test FPR/TPR between various particles. In the point source run case, we consider the ROC curves for classification between $\gamma$-rays and protons and between $\gamma$-rays and electrons. This demonstrates the combined classification power of both morphological and directional information. The equivalent curves for the diffuse case highlight solely the morphological classification power. The proton versus ($\gamma$+e) curve demonstrates the difficulty  of distinguishing between $\gamma$-rays and electrons.
  }
  \label{fig:ROC2}
\end{figure}

 The initial training took approximately 3 days on a NVidia 1080Ti GPU for each ConvLSTM2D network, meaning the training time for the results presented in Figures \ref{fig:trainlog}, \ref{fig:ROC} and \ref{fig:ROC2} was around a month. Each network for each method (A through D) and each of the two datasets (point source and diffuse) was trained from scratch independently, so each sub-figure in Figure \ref{fig:ROC} represents the test results for a uniquely trained classifier (the results from which are re\-plotted in Figure \ref{fig:ROC2}). We do not mix classifiers trained on different datasets, so classifiers trained on point source data are tested on (previously unseen) point source data, and not diffuse data. As such, and given the fact that there is no preferential direction in any of diffuse data CORSIKA/sim\_telarray events (as seen in Table \ref{table:Datasetparams}), the diffuse data results demonstrate classification power based solely upon shower morphology. Training accuracies as a function of epoch for this initial training are shown in Figure \ref{fig:trainlog}. This training was a significant computational task, and represents the limit of what can be performed with a single GPU. It should also be noted that in this particular use case, the `Compute Capacity' (NVidia's measure of a GPU's specifications) of the GPU is an important attribute due to the presence of RNN features in our network. Additionally, having a large amount of VRAM allowed for the processing of large datasets.
 

\begin{table}[t]
    \centering
    \resizebox{\textwidth}{!}{
    \begin{tabular}{c|c|c}
    \textbf{Method} & \textbf{Point Source Run Classification Rate (Hz)}& \textbf{Diffuse Run Classification Rate (Hz)}\\
    \hline
    A & 308 & 296\\
    B & 240 & 231\\
    C & 339 & 320\\
    D & 308 & 336\\

    \end{tabular}  
    }
    \caption{Classification rates (the number of classifications per second) based on testing against $\sim$100,000 events for each method using 4 telescopes. Note that due to the LSTM features in our network this is substantially slower than classifications can be performed for Mono-telescope analysis ($\sim$1 kHz). This is also dependent on the GPU used (a NVidia 1080Ti in this case), and the data loader has not been fully optimised for speed. The test system had 32GB of Random Access Memory and operated at a clock speed of 1866 MHz.}
    \label{table:speed}
\end{table}
 \begin{figure}[t]
  \centering
  \includegraphics[width=\textwidth]{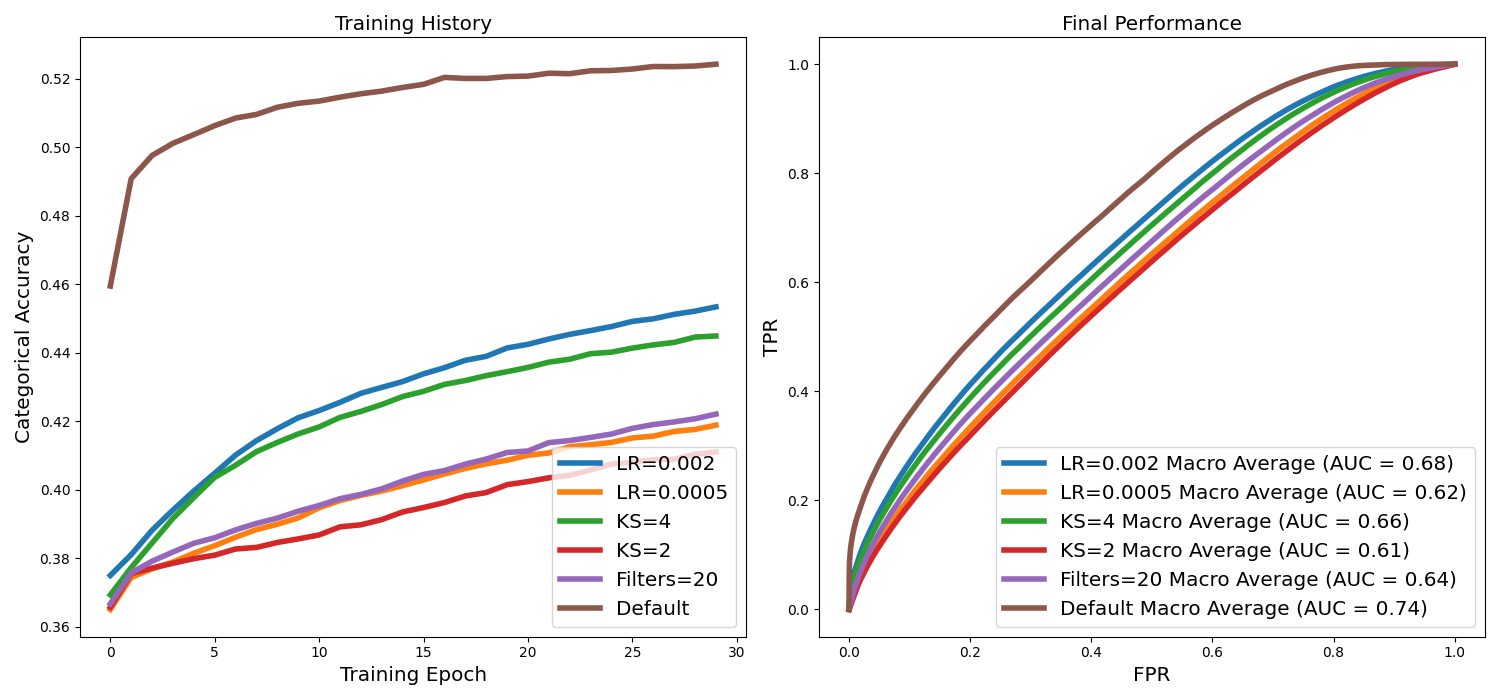}
  \caption{This plot shows the effect of modifying the hyperparameters for training Method B on the diffuse data. Left: The categorical accuracy as a function of epoch for the modified hyperparameter training runs. Right: The associated test ROC curves.
  }
  \label{fig:modhyp}
\end{figure}

Determining the source of classification power with CNN-type classifiers is non-trivial. Whilst gradient and perturbation based attribution techniques have been developed for standard single-image CNN classifiers (which can produce heat maps showing pixels relevant for classification) \cite{deepexplain}, open source packages for this do not currently extend support for such methods to the multiple image ConvLSTM2D case. As such, we are forced to interpret performance of the classifiers on different datasets in order to infer the information being used for classification. In our case, the results for the point source run and methods A and B achieve high performance metrics, but it appears as if this is as the classifier is largely using the directional information about the shower to perform classification, as opposed to using morphological features in the images. This can be seen in Figures \ref{fig:ROC} and \ref{fig:ROC2}, as the classifier differentiates between $\gamma$-rays and other classes easily in the point source case for methods A and B, but not between the diffuse protons and electrons. However, it is noteworthy that the ConvLSTM2D network performs this task so well and that methods A and B perform this in a superior way to methods C and D. This bodes well for future studies into directional reconstruction using CNN-based methods, as well as demonstrating the merit for multi-task deep learning studies that attempt to perform the tasks of event classification and directional reconstruction simultaneously (such as \cite{jacquemont}). This knowledge of the strong directional sensitivity of CNN-type methods can be used to inform the design of future CNN-based analysis pipelines for CTA.

In the diffuse case (where all three event classes come from an even spread of directions), methods C and D overtrain and cannot distinguish between any of the event classes. However the timing based methods A and B appear able to perform good event classification based on shower morphology in Figure \ref{fig:ROC2} between the (combined) $\gamma$-ray and electron showers versus protons (that could be optimised further), however they cannot distinguish between the $\gamma$-ray and electron induced showers. The classification rates for the four different methods are shown in Table \ref{table:speed}, despite using many more pixel maps method B classifies at $\sim 70\%$ the rate of the other methods. However, although including much more information than method A, method B doesn't appear to offer any significant improvement. This is likely as the timing information is likely to be the key component in the classification, though the marginally increased number of trainable parameters (as seen in Table \ref{table:methods}) for method B might disadvantage it slightly on a differential test basis.

In order to test the effect of varying hyperparameters, we ran additional training runs for Method B on the diffuse data. These runs were identical to the previous training runs for method B bar a single change each from the `default' hyperparameters shown in Table \ref{table:hyperparams} and Figure \ref{fig:model}. One run each had doubled and halved the Learning Rate (LR) from the default Adadelta value of 0.001, one run had all ConvLSTM2D kernel filter sizes (KS) set to 2x2, one with all filter sizes set to 4x4 and a final run with the number of filters per layer set to 20 rather than 30 (this value could not be increased due to VRAM constraints). The training accuracy curves (all showing reasonable convergence) and test results from this are shown in Figure \ref{fig:modhyp}. The `default' hyperparameter setup outperforms the alternatives, however this is to be expected as changing the \textit{Adadelta} optimiser parameters is not recommended by the developers of \textit{Keras} \cite{Keras}, and the 3x3 filter size option also provided the good results in the original ConvLSTM2D paper \cite{shi}. Whilst this gives us some idea of the spread in performance due to hyperparameter selection, this is not a substitute for an exhaustive hyperparameter space search, the optimal combinations of which are often not human-interpretable. The uncertainty associated with individual event classifications is also not trivial to obtain with CNN-type methods \cite{mike}\cite{gal2015}, and this is likely to be a necessary area of further development for these analysis techniques.

 \begin{figure}[t]
  \centering
  \includegraphics[width=\textwidth]{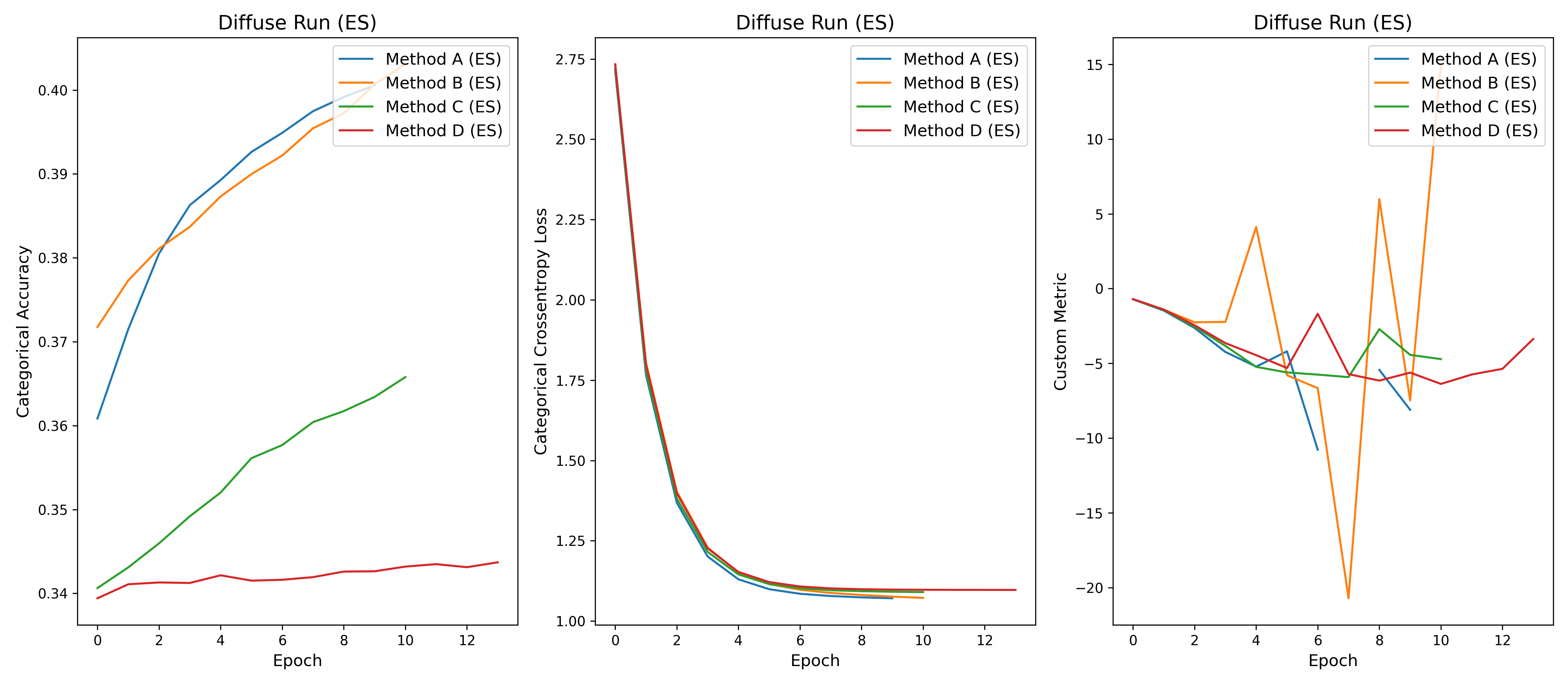}
  \caption{Training total loss, categorical accuracy, and custom metric (M) values for the four ES runs, all on the diffuse data. The custom metric curve for method A hits an infinite value on epoch 7.
  }
  \label{fig:trainlogES}
\end{figure}

\begin{figure}
  \centering
  \includegraphics[width=\textwidth]{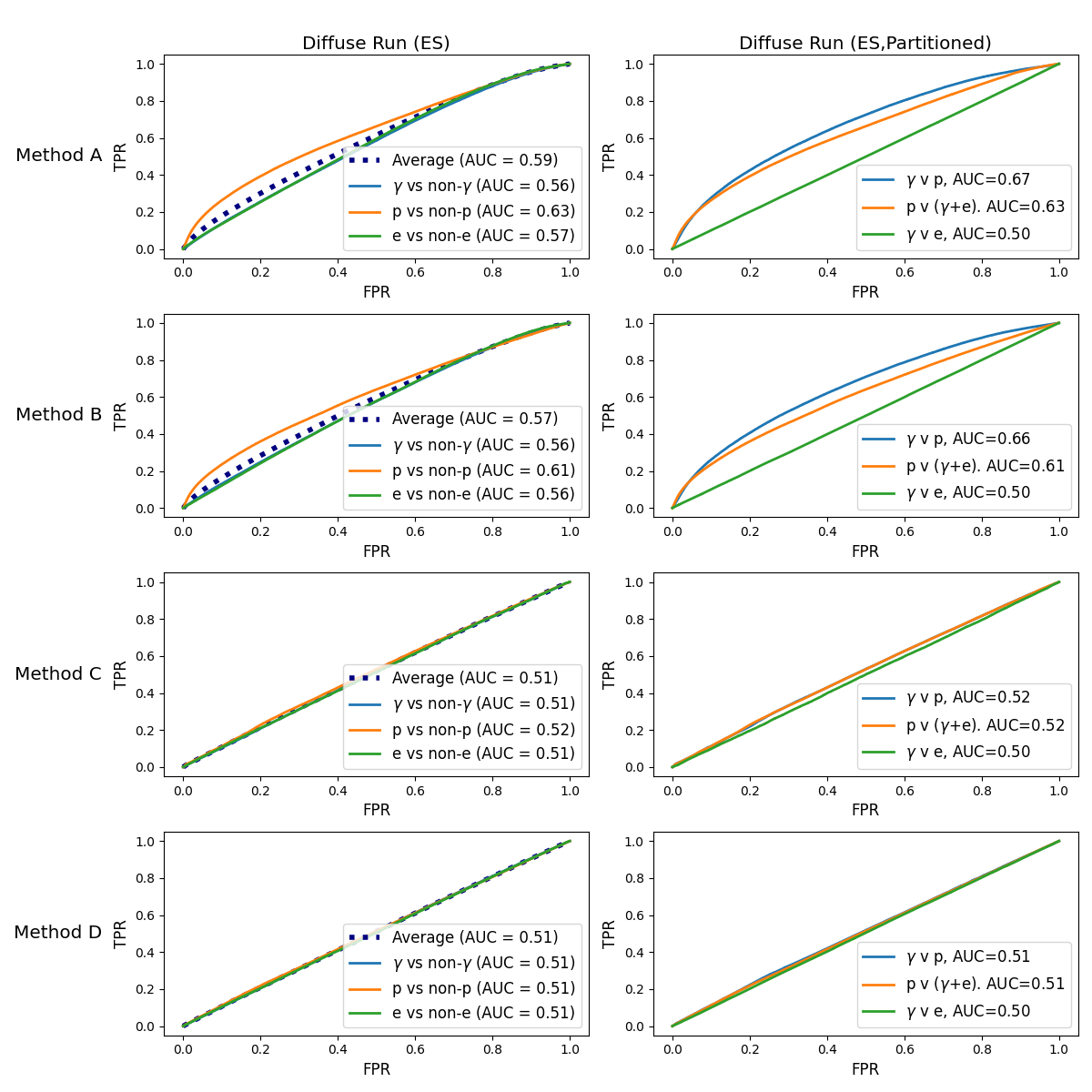}
  \caption{This figure shows the final results from the ES runs, which are only trained and tested on diffuse data. Left: The results presented in the standard machine learning one versus all approach as in Figure \ref{fig:ROC}. Right: The same ES results shown in the astrophysical context of Figure \ref{fig:ROC2}.
  }
  \label{fig:ROC3}
\end{figure}

In order to investigate the influence of overtraining of methods C and D in the initial diffuse runs (as seen in Figure \ref{fig:trainlog}) with the original hyperparameter configuration from Table \ref{table:hyperparams} and Figure \ref{fig:model}, we ran additional training and testing runs on this diffuse data using this `default' configuration along with two early stopping criteria (that we reference by the abbreviation ES). Early stopping in \textit{Keras} allows for the stopping of training when there is an inversion in a chosen metric function, and for the restoration of the earlier optimal model weights obtained during training. In the case of preventing the overtraining scenario we observed for method D (where there's a clear inversion in the loss at epoch 6), the choice of this early stopping metric is the total loss function used during training. However, choosing such a metric to prevent the overtraining scenario we observed for Method C whereby there is undesired, runaway acceleration in the training data loss is more complex. As such, we implemented a custom second metric (M) of the form
\begin{equation}
    M=\frac{1}{0.7-0.8*L}
\end{equation}
where L is the total loss shown in for example \ref{fig:trainlog}. This L includes both the categorical cross-entropy loss and the regularization penalties which we explicitly sum over the first two ConvLSTM2D layers. We intentionally designed this metric as it has a turning point (that will trigger the \textit{Keras} early stopping) around the epoch that method C appeared to start overtraining in the original training runs. For both metrics we choose a \textit{patience} parameter of 3, representing the number of epochs to tolerate the inversion of the metrics.

The training curves and results from these ES runs can be seen in Figures \ref{fig:trainlogES} and \ref{fig:ROC3}. During the ES training for Method A, the early stopping was triggered on epoch 10, for Method B and C on epoch 11, and Method D on epoch 14. We still observe superior performance (and some ability to perform gamma/hadron separation using morphological information) for the timing based methods A and B, despite this early stopping procedure (which was human-engineered for Method C). As such, we believe the superior performance of methods A and B in the diffuse case cannot be solely attributed to a bias from the non-convergence of the initial training for methods C and D, and is a result of the charge-only dataset used for methods C and D rather than an effect due to the design or optimisation of the classifier network.

The charge-only methods \cite{Shilon} (methods C and D) perform poorly in our analysis in both the point source and diffuse cases (regardless of whether early stopping is used or not); this isn't to say that such charge-only methods could not function well given additional hyperparameter optimisation and image cleaning. We explained our motivation for not performing such hyperparameter optimisation in section \ref{Methods}. We seek simply to demonstrate the effectiveness of our timing techniques on a differential basis, which is confirmed by our results. As such, the absolute ROC, AUC, loss and accuracy values here should not be considered to be optimal for any of the four methods. It should also be noted that the realistic presence of electron events in the training datasets likely makes training the networks in the diffuse case significantly harder than straightforward $\gamma$-hadron separation given their similarity to $\gamma$-ray events, and this is likely part of the reason these two methods perform poorly in the diffuse case. Part of our reasoning behind these conclusions is that all four methods are affected to roughly the same extent ($\sim$0.2 AUC) by the change between (non-ES) diffuse and point source runs. Small variations in the accuracy values between the different methods could be statistical; performing multiple trials with different random seeds might help to confirm this, however this would be highly computationally intensive.

Performing a fair comparison between existing BDT techniques and CNN-based methods is an involved problem, especially as CTA reference analyses have yet to be established. Whilst an initial comparison can be found in the Shilon et. al work \cite{Shilon}, currently unsolved questions about how to robustly perform such comparisons include the quantification of the computational cost, the management of the sensitivity versus robustness trade-off, and the correct way to co-optimise the classifiers. Whilst BDTs may be trivial to optimise, the same cannot be said for CNN-type methods, and this ultimately this might limit their practicability. This will be the subject of a future CTA consortium technical publication, and as a result we do not attempt such a comparison here.

\section{Conclusions} \label{Conclusions}
We have investigated four methods of applying new deep learning methods to the task of classifying events for IACTs, finding that the use of timing information is an viable method of event discrimination against simulated hadronic air showers. Whilst not a full sensitivity analysis, we feel our results demonstrate the value in including timing information in future CTA deep learning analyses. This became possible thanks to the significant technical innovation that has gone into the design of CHEC and other similar modern Cherenkov cameras. However, we believe our work demonstrates that there is little prospect of CNN-based methods being used to filter out electron induced air showers, and so in future IACT deep learning studies (in the SST energy range) electrons should be treated as an irreducible background. That said, future IACT instruments may provide even more detailed images of EAS that might make electron event rejection feasible.

In considering future prospects, we  stress that our ConvLSTM2D networks have been trained entirely with simulated data.  There are potentially serious issues to be considered when applying these CNN-based methods trained with simulations to real data. Variations in night sky background levels, dead pixels or malfunction of front end electronics, bright stars in the field of view, cloud cover and the ageing of telescopes could all create biases in a CNN-type classifier response, and not all of these effects are trivial to simulate. As a result, achieving high performance metrics for CNN-based classifiers on simulations does not (on its own) correspond to an effective increase in IACT sensitivity on-sky. It appears clear that new simulation and analysis approaches are necessary in order to attack this problem \cite{rws}. However, there is a wide body of computer science literature on this problem of `Unsupervised Domain Adaptation' which could provide solutions. Key approaches include methods based on Generative Adversarial Networks (GANs) \cite{cycada} and data augmentation.

There are astrophysical implications as well that require detailed study. As the Shilon et al. authors correctly raise \cite{Shilon}, not every source on the $\gamma$-ray sky is equally bright. As such, cut optimisation on gamma-likeness values is frequently applied when performing event classification with BDTs or RFs, and this is not easy to replicate using a CNN-type background rejection method. We speculate that the use of Bayesian Convolutional Neural Networks \cite{mike}\cite{gal2015} may assist in providing meaningful estimates of classification uncertainty for individual events, and these might aid this task. However propagating such uncertainty through a complete IACT analysis would be non-trivial. An investigation into how to mitigate some of these issues affecting real data will be presented in a later work, though it is clear that collaboration and data sharing with existing instruments will be key in order to investigate these issues.  

\section{Acknowledgements}
This work used the European Grid Infrastructure. We thank Luisa Arrabito and Johan Bregeon for their assistance in running our simulations and code on it. Later training runs in this paper used the \textit{Glamdring} cluster in Oxford; we thank Jonathan Patterson for his work in upgrading and maintaining this facility. SS acknowledges an STFC Ph.D. studentship. GC, JW and TA acknowledge support from STFC grants ST/S002618/1 and ST/M00757X/1. GC acknowledges support from Exeter College, Oxford. This work has gone through internal review by the CTA Consortium, who we thank for allowing us to use CTA simulation models. We gratefully acknowledge financial support from the agencies and organisations listed here: \url{http://www.cta-observatory.org/consortium\_acknowledgments}.

\appendix

\section{Further Basic Machine Learning Definitions}
\label{MLdefs}
In this appendix, we provide additional machine learning definitions that may assist the reader. In order to train a neural network, one needs to define a loss function for the training algorithm to optimise against. For multi-class classification the most appropriate choice is a \textbf{Categorical Cross-Entropy Loss Function (CE)}. This is defined \cite{Keras} as :
\begin{equation}
    \textrm{CE}=-\sum_i^N \ln \left( \frac{e^{t_{i}}}{\sum_j^3 e^{s_{ij}}} \right)
\end{equation}

where there are N events, $s_{ij}$  is the CNN score for each of the possible event classes for the event $i$ and $t_{i}$ is the CNN score for the target (true) class for that event. This equation relies on the CNN input labels being `one-hot' encoded such that the label for class $0$ is represented as $[1,0,...]$, class $1$ as $[0,1,...]$ so on \cite{fb}. In practice, during our training, there is an additional regularization penalty added to the total loss used due to the L2 regularization we use in the first two ConvLSTM2D layers. This regularization is designed to prevent network weights from taking extreme values.

\begin{table}[ht]
    \centering
    \resizebox{0.7\textwidth}{!}{
    \begin{tabular}{c|c}
    \textbf{Term} & \textbf{Definition}\\
    \hline
    \textbf{True Positive (TP)} & True Label Positive and Classification Positive\\
    \textbf{True Negative (TN)} & True Label Negative and Classification Negative\\
    \textbf{False Positive (FP)} & True Label Negative and Classified Positive\\
    \textbf{False Negative (FN)}& True Label Positive and Classified Negative\\
    \end{tabular}  
    }
    \caption{Definitions of Event Classifications for binary classification, taken from \cite{fawcett}.}
    \label{table:FPR}
\end{table}

For the purposes of evaluation of network performance, \textbf{Categorical Accuracy} \cite{Keras} is defined as the percentage of events for which the \textit{Argmax} of the one-hot-encoded predicted label is the same as the \textit{Argmax} for the one-hot-encoded true label (the definition of binary accuracy is similar for 2-class classification \cite{Keras}). Given the definitions for binary classification in Table \ref{table:FPR}, the \textbf{True Positive Rate (TPR)} and \textbf{False Negative Rate (FNR)} are defined by \cite{fawcett}:
\begin{equation}
    TPR=\sum_{Events}\frac{TP}{TP+FN}=1-FNR
\end{equation}
and the \textbf{True Negative Rate (TNR)} and \textbf{False Positive Rate (FPR)} by
\begin{equation}
    TNR=\sum_{Events}\frac{TN}{TN+FP}=1-FPR.
\end{equation} For multi-class ($>2$ class) classification, such as in Figure \ref{fig:ROC}, we use a one-versus-all approach where for each individual class the classifications are treated in a binary way (such as $\gamma$-ray or not-$\gamma$-ray) \cite{fawcett}\cite{scikit}. 
\bibliographystyle{elsarticle-num} 
\bibliography{main}

\end{document}